\documentclass[fleqn,10pt]{wlscirep}
\usepackage[utf8]{inputenc}
\usepackage[T1]{fontenc}

\title{Characterization of a novel automated microfiltration device for the efficient isolation and analysis of circulating tumor cells from clinical blood samples}

\author[1,*]{Juan F. Yee-de~Le\'on}
\author[1,*]{Brenda Soto-Garc\'ia}
\author[1,*]{Diana Ar\'aiz-Hern\'andez}
\author[1,2,*]{Jes\'us Rolando Delgado-Balderas}
\author[1]{Miguel A. Esparza}
\author[1]{Carlos Aguilar-Avelar}
\author[1,3]{J. D. Wong-Campos}
\author[1]{Franco Chac\'on}
\author[1]{Jos\'e Y. L\'opez-Hern\'andez}
\author[1]{A. Mauricio Gonz\'alez-Trevi\~no}
\author[1]{Jos\'e R. Yee-de~Le\'on}
\author[1]{Jorge L. Zamora-Mendoza}
\author[4,5]{Mario  M.  Alvarez}
\author[4,6]{Grissel Trujillo-de~Santiago}
\author[7]{Lauro S. G\'omez-Guerra}
\author[2]{Celia N. S\'anchez-Dom\'inguez}
\author[1,+]{Liza P. Velarde-Calvillo}
\author[1,+]{Alejandro Abarca-Blanco}

\affil[1]{Delee Corp., Mountain View, CA, 94041, USA.}
\affil[2]{Departamento de Bioqu\'imica y Medicina Molecular, Facultad de Medicina, Universidad Aut\'onoma de Nuevo Le\'on, Monterrey, 64460, Mexico.}
\affil[3]{Department of Physics, Joint Quantum Institute and Joint Center for Quantum Information and Computer Science, University of Maryland, College Park, MD, 20742, USA.}
\affil[4]{Centro de Biotecnolog\'ia-FEMSA. Escuela de Ingenier\'ia y Ciencias, Tecnologico de Monterrey, Monterrey, 64849, Mexico.}
\affil[5]{Departamento de Bioingenier\'ia, Escuela de Ingenier\'ia y Ciencias, Tecnologico de Monterrey, Monterrey, 64849, Mexico.}
\affil[6]{Departamento de Mecatr\'onica e Ingenier\'ia El\'ectrica, Escuela de Ingenier\'ia y Ciencias, Tecnologico de Monterrey, Monterrey, 64849, Mexico.}
\affil[7]{Servicio de Urolog\'ia, Hospital Universitario ``Dr. Jos\'e Eleuterio Gonz\'alez'', Universidad Aut\'onoma de Nuevo Le\'on, Monterrey, 64460, Mexico.}

\affil[+]{Correspondence and requests for materials should be addressed to L.P.V. (liza@delee.bio) or A.A (alejandro@delee.bio)}
\affil[*]{These authors contributed equally to this work}

\begin{abstract}

The detection and analysis of circulating tumor cells (CTCs) may enable a broad range of cancer-related applications, including the identification of acquired drug resistance during treatments. However, the non-scalable fabrication, prolonged sample processing times, and the lack of automation, associated with most of the technologies developed to isolate these rare cells, have impeded their transition into the clinical practice.
This work describes a novel membrane-based microfiltration device comprised of a fully automated sample processing unit and a machine-vision-enabled imaging system that allows the efficient isolation and rapid analysis of CTCs from blood. The device performance was characterized using four prostate cancer cell lines, including PC-3, VCaP, DU-145, and LNCaP, obtaining high assay reproducibility and capture efficiencies greater than 93\% after processing 7.5 mL blood samples from healthy donors, spiked with 100 cancer cells. Cancer cells remained viable after filtration due to the minimal shear stress exerted over cells during the procedure, while the identification of cancer cells by immunostaining was not affected by the number of non-specific events captured on the membrane. We were also able to identify the androgen receptor (AR) point mutation T878A from 7.5 mL blood samples spiked with 50 LNCaP cells using RT-PCR and Sanger sequencing.
Finally, CTCs were detected in 8 of 8 samples from patients diagnosed with metastatic prostate cancer (mean $\pm$ SEM = 21 $\pm$ 2.957 CTCs/mL, median  = 21 CTC/mL), thereby validating the potential clinical utility of the device. 

\end{abstract}

\begin{document}

\flushbottom
\maketitle
%
%
\thispagestyle{empty}

\section*{Introduction}

In the last two decades, circulating tumor cells (CTCs) have attracted a significant amount of attention for their potential use as a blood-based biomarker for a broad range of cancer-related clinical applications. CTCs are malignant cells that are shed from the primary and/or metastatic solid tumors and then infiltrate into the vascular and lymphatic systems; these cells play a fundamental role in the metastatic process of non-hematological cancers \cite{fidler2003, dotan2009, pantel2016}. Although the first report describing the existence of CTCs dates from 1869 \cite{ashworth1869}, the heterogeneity and the extremely low concentration of these cells in regard to the cellular components of blood, about 1-10 CTCs per $10^{9}$ blood cells, makes their capture extremely challenging \cite{yu2011, haber2014}. It was not until the recent development of technologies with the required sensitivity and reproducibility, that the possibility to perform CTC-based clinical assays started to become a reality.

To date, numerous studies have shown that CTCs can be used as a prognostic indicator of disease progression and overall survival in patients with metastatic breast, prostate and colorectal cancer \cite{budd2006, debono2008, cohen2009, miller2010}. In addition, changes in the CTC burden in patients over time have been associated with the effectiveness of the administered therapies \cite{stott2010a, lu2013,  Lorente2018}. Furthermore, the phenotypic and genotypic analysis of CTCs can enable the continuous assessment of mutations that confer therapeutic sensitivity or resistance to targeted therapies, providing information that is of paramount importance for cancer treatment personalization \cite{maheswaran2008, ignatiadis2011, Steinestel2015}. Recent studies suggest that CTCs may even have the potential to be used as a biomarker for recurrence and early cancer detection \cite{stott2010a, galizia2013, Zhang2018}. 

Many of the technologies developed to isolate CTCs from blood are based on sample enrichment methods that depend on specific antigen-antibody interactions, such as microfluidic devices functionalized with biomolecules that act as targeting ligands \cite{nagrath2007, stott2010b, kirby2012} or platforms that use micro- or nano-magnetic particles coated with specific antibodies to isolate these rare cells  \cite{allard2004, talasaz2009, harb2013}. Although these technologies have demonstrated clinical utility, a fundamental problem of these approaches is the lack of a universal surface marker that is consistently expressed by CTCs. Most of these technologies, including the CellSearch\textsuperscript{\textregistered} System, which is considered the gold standard, use EpCAM (epithelial cell adhesion molecule) antibodies to selectively trap cancer cells to the functionalized substrate/particles \cite{gabriel2016}. However, CTCs intravasate into the bloodstream by undergoing a process known as the epithelial-mesenchymal transition (EMT), in which their epithelial phenotype is downregulated, including the expression of EpCAM antigens. This fact limits the capture of CTCs subpopulations with diminished expression of this specific surface marker, thereby losing valuable information \cite{gorges2012, francart2018}. Therefore, there exists a need for technologies with different capture approaches that are independent of surface markers expressed by CTCs.

An effective alternative are microfiltration devices, which rely on the differences in size and deformability between blood cells and CTCs to capture these rare cells. These platforms have consistently proven their effectiveness at isolating a greater number of CTCs in samples from patients with different types of cancer, even capturing CTCs subtypes that no longer express EpCAM antigens, when compared with approaches based on capture antibodies \cite{lin2010, adams2015, gogoi2016}.

Various microfiltration technologies such as the commercially available ISET\textsuperscript{\textregistered} \cite{vona2000} and ScreenCell\textsuperscript{\textregistered} \cite{desitter2011} devices, use polycarbonate track-etch membranes. However, the processes used to fabricate these membranes create random pores in the plastic sheet, making these membranes inadequate for the isolation of CTCs. The porosity of these membranes cannot exceed from 2\% without overlapping between pores, while increasing the number of pores beyond this point can hinder the efficiency of separation of cells of similar size in regard to the pore size \cite{fleischer1964}. The membranes are therefore restricted to a low porosity, which can lead to cell damage (due to uneven distribution of pressure over the membrane), lower capture efficiency, reduced reproducibility, and a higher recovery of non-specific cells (which can potentially clog the membrane) \cite{zheng2007, adams2014}.

To circumvent these issues, other approaches have utilized membranes crafted through microfabrication techniques. The advantage of these fabrication processes is the capacity to produce membranes with uniform patterns, where the size, geometry, quantity, and distribution of pores can be precisely controlled; allowing higher porosities and thereby mitigating the issues aforementioned. To capture CTCs, several groups have employed microfabricated membranes made from different materials, such as parylene C \cite{zheng2007, xu2010, liu2019}, SU-8 \cite{adams2014, kang2015}, and silicon \cite{lim2012, dewit2015}. However, the fabrication processes for these membranes are complex, making their production costly and hardly scalable, limiting their introduction into clinical practice. Recently, photolithography-based electroforming, a technique that enables the production of highly precise metal structures, has been utilized to fabricate filters used for the isolation of cellular subpopulations, including CTCs from blood samples \cite{hosokawa2010, hosokawa2013, yusa2014}. In contrast with the ones created with other manufacturing methods, these membranes could be mass produced at a significantly lower cost. 

Nevertheless, despite of the membrane used, the microfiltration technologies developed so far, are not capable of processing, preparing, and analyzing the captured cells in an automated way \cite{vona2000, zheng2007, hosokawa2010, lin2010, xu2010, desitter2011, lim2012, adams2014, yusa2014, adams2015, kang2015, dewit2015, liu2019}; easily leading to human error and cell loss due to the manual steps that must be performed. Additionally, performing a proper enumeration of the captured CTCs requires optical analysis with highly specialized microscopes that may not be available in all clinical settings.

In order to standardize the use of CTCs as a biomarker in clinical practice, it is fundamental to acknowledge the necessity of developing platforms with high recovery rates and the capacity of carrying out the entire sample workflow, without human intervention. In this manuscript, we present a novel automated microfiltration device that utilizes electroformed nickel membranes and a machine-vision-enabled microscope specifically designed for the efficient isolation and analysis of CTCs. The performance of this platform is characterized using blood samples from healthy donors spiked with prostate cancer cells, and the results obtained after processing samples from patients with diagnosed meta\-static prostate cancer are also discussed in this work.

\section*{Methods}

\subsubsection*{Blood extraction} 

Blood samples from healthy male donors and patients with diagnosed metastatic prostate cancer were provided by the urology service of the ``Dr. Jos\'e Eleuterio Gonz\'alez'' University Hospital, according to the protocol approved by their Institutional Review Board with number UR16-0007. Prior to blood extraction, informed consent was obtained from healthy donors and cancer patients. Samples were collected in 6 mL BD Vacutainer\textsuperscript{\textregistered} K\textsubscript{2}EDTA blood collection tubes (BD, Franklin Lakes, New Jersey, USA) and were processed through our device within 3 hours after the extraction. All procedures involving human participants were performed in accordance with the 1964 Helsinki declaration and its later amendments or comparable ethical standards.

\subsubsection*{Cell culture of cancer cell lines}

Human prostate cancer cell lines (ATCC\textsuperscript{\textregistered}, Manassas, Virginia, USA), PC-3 (CRL-1435\texttrademark), LNCaP clone FGC (CRL-1740\texttrademark), DU-145 (HTB-81\texttrademark), and VCaP (CRL-2876\texttrademark) were cultured using F-12K, RPMI-1640, EMEM, and DMEM media (ATCC\textsuperscript{\textregistered}, Manassas, Virginia, USA), respectively. All media were supplemented with 10\% fetal bovine serum (FBS) (ATCC\textsuperscript{\textregistered}, Manassas, Virginia, USA). Cultures were incubated at 37$^\circ$C in an air atmosphere of 5\% CO\textsubscript{2} and harvested from T-25 flasks using 0.25\% trypsin - 0.02\% EDTA solution (Thermo Fisher Scientific, Waltham, Massachusetts, USA) when 80\% confluence was reached. Cells were quantified with a hemocytometer and viability was assessed by trypan blue dye exclusion. A viability of at least 95\% was achieved after harvesting.

\subsubsection*{Preparation and processing of spiked blood samples for capture efficiency and purity assessment}

To evaluate device performance, PC-3, LNCaP, DU-145, or VCaP prostate cancer cells were stained prior to being spiked into blood samples from healthy donors by incubating cell cultures with 25 $\mu$M of CellTracker\texttrademark\ orange CMRA (Thermo Fisher Scientific, Waltham, Massachusetts, USA) for 45 min at 37$^\circ$C in an air atmosphere of 5\% CO\textsubscript{2}. Cells were harvested using 0.25\% trypsin - 0.02\% EDTA solution (Thermo Fisher Scientific, Waltham, Massachusetts, USA) and counted with a hemocytometer. The resulting cell suspension was serially diluted to achieve a concentration of approximately 100 cancer cells per 30 $\mu$L; that volume was then added to the blood samples. Subsequently, spiked blood samples were diluted with a 0.3\% formaldehyde - 0.15\% plu\-ronic F68 solution in PBS to a 1:2 v/v ratio and incubated for 10 min at room temperature before being processed by the device, at either 2 or 3 mL/min flow rates, using membranes with a 7, 8, or 9 $\mu$m pore size. Once the samples were filtered, 0.1\% pluronic F68 solution in PBS was flowed to wash out blood cells remaining on the membrane, followed by fixation and nuclear staining which were carried out by incubating the membrane for 10 min after passing 1 mL of 4\% formaldehyde and 1 $\mu$g/mL Hoechst 33342 solutions at 500 $\mu$L/min, respectively. At the end of each incubation, 0.1\% pluronic F68 solution in PBS was used to wash the remnants of the fixative and nuclear staining dye. Finally, the holder was disassembled and the membrane was mounted on a microscope slide using Fluoromount-G\texttrademark\ (Thermo Fisher Scientific, Waltham, Massachusetts, USA) for its subsequent analysis by fluorescence microscopy. 

To properly estimate the number of tumor cells spiked into blood samples, equal volumes of the cancer cell suspension were added to 10 wells of a 96-well plate. The capture efficiency of the microfiltration device was determined by comparing the number of cancer cells trapped by the membrane against the average number of cells counted on the wells. To calculate purity, the total number of tumor cells was divided by the total number of nucleated events along the membrane. 

\subsubsection*{Preparation and processing of spiked blood samples for viability assessment}

The LIVE/DEAD\texttrademark\ assay was used to evaluate cell viability. PC-3 cancer cells were pre-stained with CellTracker\texttrademark\ blue CMF\textsubscript{2}HC (Thermo Fisher Scientific, Waltham, Massachusetts, USA) and spiked into samples from healthy donors at a concentration of 1000 cells per 7.5 mL of blood. Spiked samples were diluted with a 0.15\% plu\-ronic F68 solution in PBS to a 1:2 v/v ratio and incubated for 10 min at room temperature before being processed by the device. Once the samples were filtered, 0.1\% pluronic F68 solution in PBS was flowed at 500 $\mu$L/min during 10 min to wash the blood cells unspecifically captured on the membrane. Afterward, 1 mL of 2 $\mu$M calcein AM - 4 $\mu$M ethidium homodimer-1 solution was passed at 500 $\mu$L/min, followed by a 45 min incubation. Then, the holder was disassembled and the membrane was mounted on a microscope slide for its analysis by fluorescence microscopy.

Blue/green fluorescent events were classified as live tumor cells, whereas blue/red events were counted as dead cancer cells. Cell viability was calculated as a percentage by dividing the total number of live tumor cells by the total number of cancer cells along the membrane.

\subsubsection*{RT-PCR analysis and sequencing}

Healthy blood samples, spiked with 15, 50, 250, 500, and 1000 LNCaP cells were processed at 2 mL/min. After filtration, 0.1\% pluronic F68 solution in PBS was flowed through the device at 500 $\mu$L/min for 10 min to wash out cellular residues before transferring the membranes to microcentrifuge tubes for nucleic acid extraction. Total RNA was isolated from the cells captured in the membrane using the AllPrep\textsuperscript{\textregistered} DNA/RNA FFPE kit (Qiagen, Venlo, Netherlands), according to the manufacturer's instructions, excluding the deparaffinization steps. Full-length cDNA was produced from total RNA by first-strand cDNA synthesis using the SuperScript IV\texttrademark\ First-Strand Synthesis System (Thermo Fisher Scientific, Waltham, Massachusetts, USA), following the manufacturer's instructions. A 599 base pair coding region of the ligand binding domain (LBD) of the androgen receptor (AR) was amplified using the Platinum SuperFi PCR Master Mix (Thermo Fisher Scientific, Waltham, Massachusetts, USA) and the following primer pairs: forward 5$^\prime$-CCAATGTCAACTCCAGGATGCTCTAC-3$^\prime$, and reverse 5$^\prime$-AATTCCCCAAGGCACTGCAGA-3$^\prime$.
PCR amplicons were purified using the PureLink\textsuperscript{\textregistered} Quick Gel Extraction kit, for further Sanger sequencing. Sequencing electropherograms were compared to the reference sequence for the AR (NM\textunderscore000044.4) to identify the missense mutation T878A.

\subsubsection*{Preparation, processing and on-membrane immunofluorescence staining of blood samples from patients with metastatic prostate cancer}

Blood samples of 7.5 mL from patients diagnosed with metastatic prostate cancer were diluted with a 0.3\% formaldehyde - 0.15\% pluronic F68 solution in PBS to a 1:2 v/v ratio and incubated for 10 min at room temperature before being processed through the device at a flow rate of 2 mL/min. Once the samples were filtered, 0.1\% pluronic F68 solution in PBS was flowed to wash out the remaining blood cells captured by the membrane. Fixation was performed by flowing 1 mL of 4\% formaldehyde in PBS at 500 $\mu$L/min, followed by a 10 min incubation. Subsequently, permeabilization was carried out by passing 1 mL of 0.3\% PBST at 500 $\mu$L/min, followed by a 10 min incubation. Afterward, to prevent non-specific binding of antibodies, blocking was made by flowing 1 mL of 1\% BSA in 0.1\% PBST at 500 $\mu$L/min followed by an incubation of 30 min. Then, 500 $\mu$L of an antibody cocktail containing 8 $\mu$g/mL of alexa fluor\textsuperscript{\textregistered} 488 labeled anti-cytokeratin (pan reactive) (clone C-11),  alexa fluor\textsuperscript{\textregistered} 647 labeled anti-human CD45 (clone HI30), and biotin labeled anti-human PSMA (FOLH1) (clone LNI-17) antibodies (BioLegend, San Diego, California, USA) in 1\% BSA - 0.1\% PBST solution was flowed through at 250 $\mu$L/min followed by a 1 hour incubation. Finally, 500 $\mu$L of a mixture containing 1 $\mu$g/mL of Hoechst 33342 and 8 $\mu$g/mL of streptavidin-alexa fluor\textsuperscript{\textregistered} 568 conjugate (Thermo Fisher Scientific, Waltham, Massachusetts, USA) in 1\% BSA - 0.1\% PBST solution was flowed at 250 $\mu$L/min, followed by a 1 hour incubation. Washing steps of 5 min were carried out at 500 $\mu$L/min after finishing each incubation with the solutions described above. After fixation, and for the final washing step, 0.1\% pluronic F68 in PBS was used, while a 0.1\% PBST solution was utilized in between to wash out the residues of the remaining solutions. 

\subsubsection*{Statistical Analysis}

All the experiments with prostate cancer cell lines were performed in triplicates. These results, along with the ones obtained in clinical samples, are reported as mean $\pm$ standard error of the mean (SEM). The difference between means was determined by unpaired t-test, where a p-value less than 0.05 was considered statistically significant. 

\section*{Results}

\subsection*{Automated microfiltration device and imaging system}

We have built an automated microfiltration device and an imaging system for the efficient isolation and rapid analysis of CTCs from blood samples. Electroformed nickel membranes with square-shaped pores of 7, 8, and 9 $\mu$m size were employed for processing blood samples; all of them with a 17 $\mu$m pore spacing. 

\begin{figure*}
\centering
\includegraphics[width=.85\linewidth]{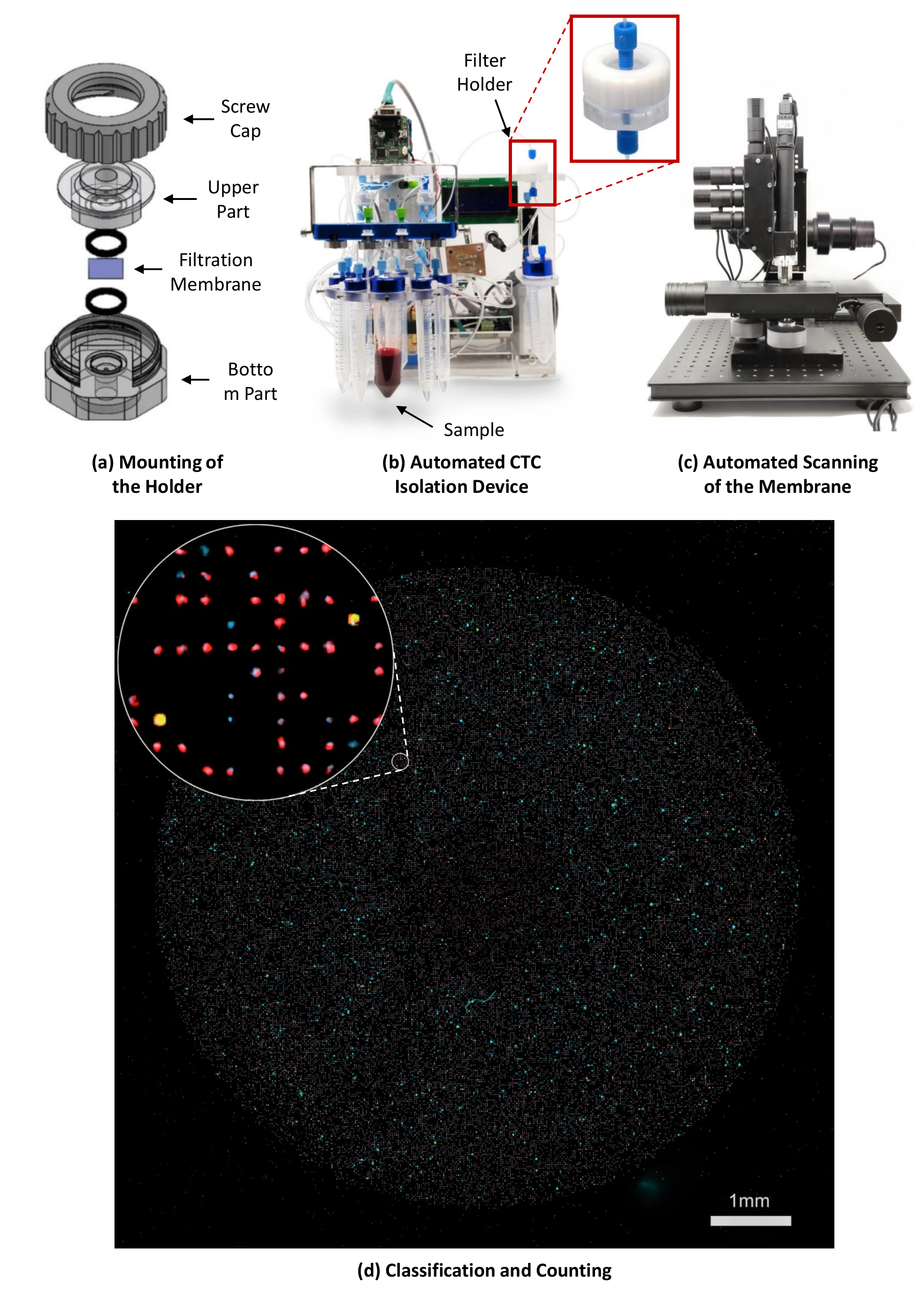}
\caption {Sample processing and analysis workflow. (a) Schematic showing the parts that composed the holder where the membrane is placed. (b) A blood sample is automatically processed through the microfiltration device, followed by immunostaining steps. The membrane is mounted on a microscope slide and images are acquired using the integrated imaging system (c). (d) Fluorescent events are automatically classified and counted by a machine-vision algorithm implemented in the imaging system. Scale bar: 1 mm.}
\label{fig:CTCPlatform}
\end{figure*}

To prevent leakage during sample processing, each membrane was placed on a custom-made PMMA holder fabricated with a computer numerical control (CNC) micro-milling machine (MDA Precision, Morgan Hill, California, USA). This holder is constituted by an upper part, bottom  part, and screw cap, as shown in Fig. \ref{fig:CTCPlatform}-a. The upper and bottom parts have a 1/4''- 28 threaded flat-bottom port that allows an easy connection to external components using standard microfluidic fittings. Both parts, also have a 0.9 mm microchannel that disembogue into a microchamber of 130 $\mu$l, designed to minimize the dead volume of the holder and ensure an appropriate distribution of flow throughout the processing area. The membrane is sandwiched between the upper and bottom parts of the holder along with two O-rings, and the setup is fastened by screwing in the cap to the thread milled in the holder's bottom part. The sample processing area, which is the surface of the membrane that is in contact with the sample, is delimited by the holder and has a 9 mm diameter. This translates to an approximate number of 110440, 101790, and 94110 pores for the 7, 8, and 9 $\mu$m pore size membranes, respectively, corresponding to a porosity of 8.50\%, 10.24\%, and 11.98\% each.

The holder is connected to a fully automated flow control unit (Fig. \ref{fig:CTCPlatform}-b), which consists of a diaphragm compressor as pressure source, an electronic proportional valve that controls the pressure applied to the sample and reagent reservoirs, an electronic rotary valve to select from which reservoir liquid is expelled, and a flow sensor that provides the necessary feedback to precisely compute the pressure needed in the selected reservoir to maintain the desired flow rate. The entire system is commanded by a microcontroller that follows user-defined protocols, which can be programmed using a graphical user interface in a personal computer. This platform is able to process blood samples and to perform on-filter fixation and immunostaining of the captured cells automatically, without requiring disassembly of the holder. 

After processing the sample, the membrane is mounted on a microscope slide and images of the entire membrane are acquired with our imaging system comprised of a four-channel fluorescence microscope with a motorized stage and an autofocus routine that allows the scanning of the membrane with high precision (Fig. \ref{fig:CTCPlatform}-c), and a machine vision algorithm that automatically counts the fluorescent events categorized as CTCs (Fig. \ref{fig:CTCPlatform}-d). Moreover, further corroboration of classified cells can be performed by a qualified technician through a software interface that allows the individual visualization of cells that were classified as CTCs. A detailed description and validation of the imaging system performance can be found in Aguilar-Avelar \textit{et al.}\cite{Aguilar2019} 

In addition, if molecular analyses are required, the membrane can be collected in a microtube for nucleic acid extraction using commercially available kits.

\subsection*{Capture efficiency and purity assessment} 

To characterize the performance of our microfiltration device, PC-3 cancer cells were pre-stained with CellTracker\texttrademark\ orange CMRA and spiked into blood samples from healthy donors at a concentration of 100 cells per 7.5 mL. The samples were diluted, prefixed, and processed through the device, followed by on-membrane fixation and nuclear staining, as described in the Methods section. Subsequently, an image of the entire membrane was acquired with our imaging system, and fluorescent events were automatically classified and enumerated. 

Capture efficiency and purity were evaluated using membranes with pore sizes of 7, 8, and 9 $\mu$m and flow rates of 2 and 3 mL/min. Irrespective of the membrane and flow rate tested, spiked blood samples were processed without seeing any signs of coagulation. 

\begin{figure*}[t!] 
\centering
\includegraphics[width=1\linewidth]{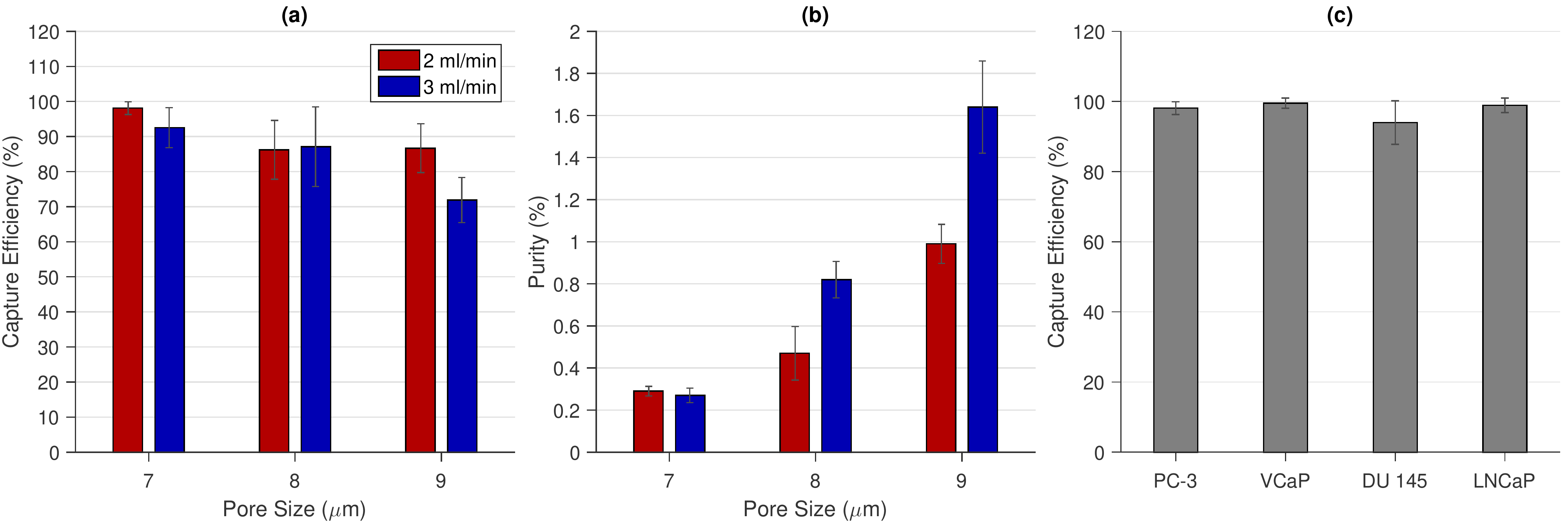}
\caption{(a) Capture efficiencies and (b) purities obtained after filtering 7.5 mL blood samples spiked with 100 PC-3 cells with membranes of pore sizes of 7, 8, and 9 $\mu$m at flow rates of 2 and 3 mL/min. (c) Capture efficiencies obtained after filtering 7.5 mL blood samples spiked with 100 PC-3, VCaP, DU-145, and LNCaP cells using membranes with a pore size of 7 $\mu$m at a flow rate of 2 mL/min. Error bars represent the standard error of the mean ($n=3$).}
\label{fig:CE_PU}
\end{figure*}

The pore size of the membrane and the flow rate at which the samples were processed, demonstrated to be directly related to the device capture efficiency, as depicted in Fig. \ref{fig:CE_PU}-a. When utilizing membranes with a 7 $\mu$m pore size and processing samples at a flow rate of 2 mL/min, an average capture efficiency of 98.09\% $\pm$ 1.82\% was obtained, dropping to 93\% $\pm$ 5.69\% when the flow rate was increased to 3 mL/min. When samples were processed through membranes with a 8 $\mu$m pore size, the capture efficiency was maintained above 85\%, regardless of the flow rate established. Similar values were observed while using membranes with a 9 $\mu$m pore size after filtering samples at 2 mL/min, but the value plummeted to 72\% $\pm$ 6.45\% when the flow rate was set to 3 mL/min. 

Overall, the rate of contaminant events decreased as the pore size of the membranes was increased and samples were processed at higher flow rates. An average of 745 $\pm$ 65 nucleated events were captured when samples were filtered through membranes of 9 $\mu$m pore size under flow rates of 3 mL/min, in comparison to the 4144 $\pm$ 315 nucleated cells captured when using membranes of 7 $\mu$m pore size and samples were processed at flow rates of 2 mL/min. Considering the number of tumor cells captured after processing spiked samples, these results corresponded to an average purity of 1.64\% $\pm$ 0.22\% and 0.29\% $\pm$ 0.02\%, respectively, as shown in Fig. \ref{fig:CE_PU}-b. 

\begin{figure*}[t!] 
\centering
\includegraphics[width=1.0\linewidth]{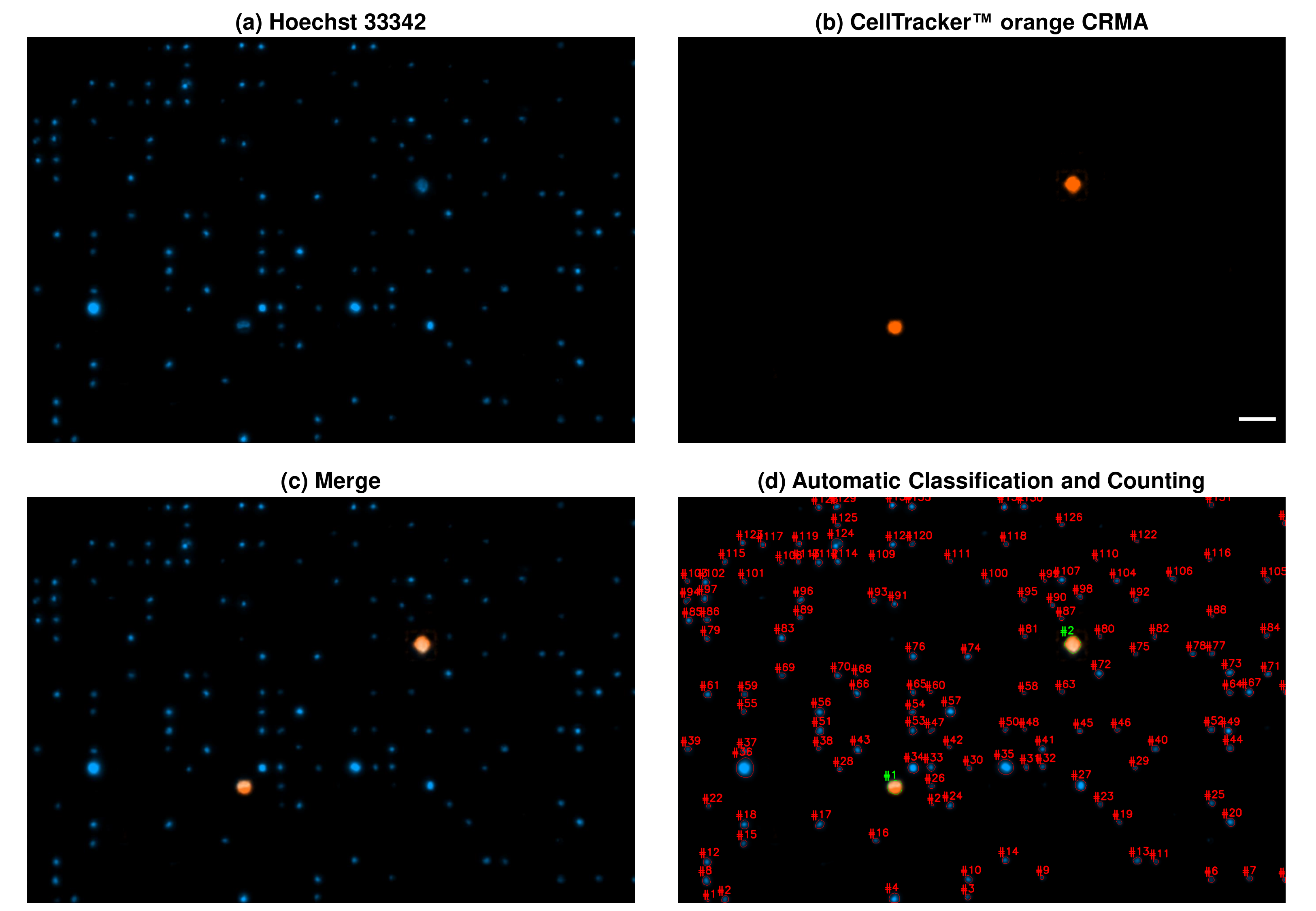}
\caption{Representative micrograph of cells captured after filtering a 7.5 mL blood sample spiked with 100 pre-stained PC-3 cells; (a) Hoechst 33342, (b) CellTracker\texttrademark\ orange CMRA, (c) Merge. (d) Automated classification and counting of fluorescent events. Scale bar: 50 $\mu$m.}
\label{fig:TwoColorSnap}
\end{figure*}

Based on the high capture efficiency obtained and despite the number of recovered non-specific events, the following experiments were carried out with membranes of 7 $\mu$m of pore size and samples being processed at a flow rate of 2 mL/min. Fig. \ref{fig:TwoColorSnap} shows a representative micrograph of the captured cells after processing a spiked blood sample under this set of parameters, as well as the results obtained after performing the automatic classification and counting.

\subsection*{Capture efficiency using different prostate cancer cell lines}

Besides the characterization made with PC-3 cells, the performance of the microfiltration device was also assessed using VCaP, DU-145, and LNCaP prostate cancer cell lines. Similarly, 7.5 mL blood samples from healthy donors were spiked with 100 pre-stained cells and processed through the device as described in the Experimental section. Capture efficiencies of 99.51\% $\pm$ 1.47\%, 94\% $\pm$ 3.60\%, and 98.92\% $\pm$ 2.06\% were obtained for VCaP, DU-145, and LNCaP, respectively, as seen in Fig. \ref{fig:CE_PU}-c. The number of contaminant events was maintained at similar levels to those described for the PC-3 cell line experiments.

The relative standard deviation (RSD), obtained from the experiments carried out with the four different cell lines, ranged between 1.93 and 6.65\%. Altogether, the results suggest that our microfiltration device has a high capture efficiency and assay reproducibility when processing spiked samples. 

\subsection*{Viability assessment}

The LIVE/DEAD\texttrademark assay was used to evaluate cell viability. To identify tumor cells from the background of blood cells captured in the membrane, PC-3 cancer cells were pre-stained with CellTracker\texttrademark\ blue CMF\textsubscript{2}HC before being spiked into the blood samples from healthy donors. Tumor cell viability was calculated by quantifying the number of blue/green (viable) and blue/red (dead) fluorescent events along the membrane. An average viability of 90.3\% $\pm$ 0.88\% was achieved after filtering spiked blood samples, in comparison to the 96\% $\pm$ 1.53\% observed in the tumor cell suspension, demonstrating that the shear stress exerted on cells is minimal, leading to high recovery rates of viable cells.

\subsection*{Molecular analysis}

\begin{figure}
\centering
\includegraphics[width=1\linewidth]{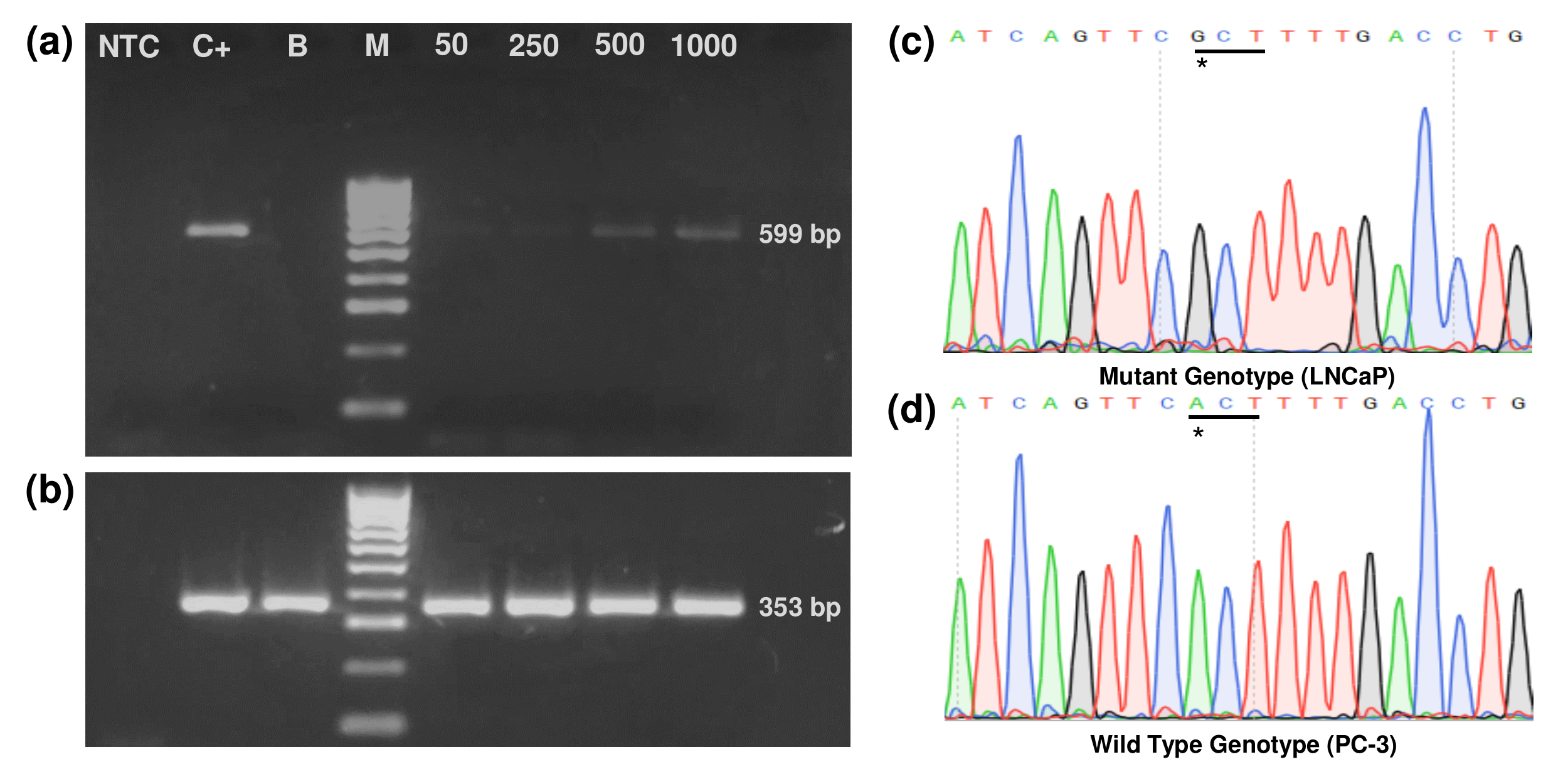}
\caption{(a) RT-PCR analysis of the AR transcript. NTC: Negative control; C+: Positive control; B: Non-spiked blood sample; M: Molecular weight marker; 50: Blood sample spiked with 50 LNCaP cells; 250: Blood sample spiked with 250 LNCaP cells; 500: Blood sample spiked with 500 LNCaP cells; 1000: Blood sample spiked with 1000 LNCaP cells. (b) ACTB was used as a control to assess cDNA synthesis. Images of the DNA stained gels were acquired using an exposure time of 330 ms and cropped to provide clarity, the uncropped images of the gels are included in Supplementary Fig. 1. The AR point mutation T878A was identified by comparing the electropherogram obtained after processing a blood sample spiked with 50 LNCaP cells (c) with the one acquired from a suspension containing only PC-3 cells (d). The underline denotes the nucleotides comprising the AR 878 codon and the asterisks indicate the nucleotide that is switched.}
\label{fig:molecular}
\end{figure}

To demonstrate that our device is compatible with standard molecular analysis, 15, 50, 250, 500, and 1000 LNCaP cells were spiked into 7.5 mL of blood samples from healthy donors and processed through the device for the identification of AR transcripts by RT-PCR. AR gene expression was consistently confirmed by agarose gel electrophoresis in samples spiked with as low as 50 cells, which is equivalent to approximately 7 cells per mL, while non-spiked blood samples were negative, as shown in Fig. \ref{fig:molecular}-a and \ref{fig:molecular}-b. In addition, the presence of the AR point mutation T878A, harbored by the LNCaP cell line, was successfully identified by Sanger sequencing after processing a blood sample spiked with 50 LNCaP cells and comparing the results with the ones derived from analyzing a sample containing only PC-3 cells, which possess the wild-type genotype, as depicted in Fig. \ref{fig:molecular}-c and \ref{fig:molecular}-d. 

The PCR products obtained from the membranes used to filter samples were successfully amplified by hemi-nested PCR (hn-PCR) when spiked with 15 cells, which is equivalent to approximately 2 cells per mL, enabling the detection of AR transcripts in two of the three experiments done with this concentration. This suggested that a lower limit of detection could be achievable with the use of more sensitive techniques, such as qRT-PCR or digital PCR (dPCR). No DNA bands were seen after reamplification of PCR reactions obtained from the membranes used to filter non-spiked blood samples.

\subsection*{Detection of CTCs from patients with metastatic prostate cancer}

To demonstrate that our platform can isolate CTCs from clinical samples, we used our device to process 8 samples of patients diagnosed with metastatic prostate cancer (age range: 56-80), who were under a therapeutic regimen, and 8 samples from healthy male controls (age range: 30-54), followed by on-membrane immunostaining as described in the protocol detailed in the Methods section. The typical definition of CTC involves a nucleated cell that expresses epithelial proteins, such as cytokeratins 8, 18, and/or 19, while being negative for the leukocyte-specific antigen CD-45. We have incorporated the prostate specific membrane antigen (PSMA) into this definition; PSMA is a transmembrane protein that is usually overexpressed in malignant prostate cancer cells and its expression increases as the disease progresses \cite{schwarzenboeck2017, boustani2018}. Nucleated cells with the phenotypes CK+/PSMA+/CD45-, CK+/PSMA-/CD45-, and CK-/PSMA+/CD45- were enumerated as CTCs. 

\begin{figure}[t!]
\centering
\includegraphics[width=0.5\linewidth]{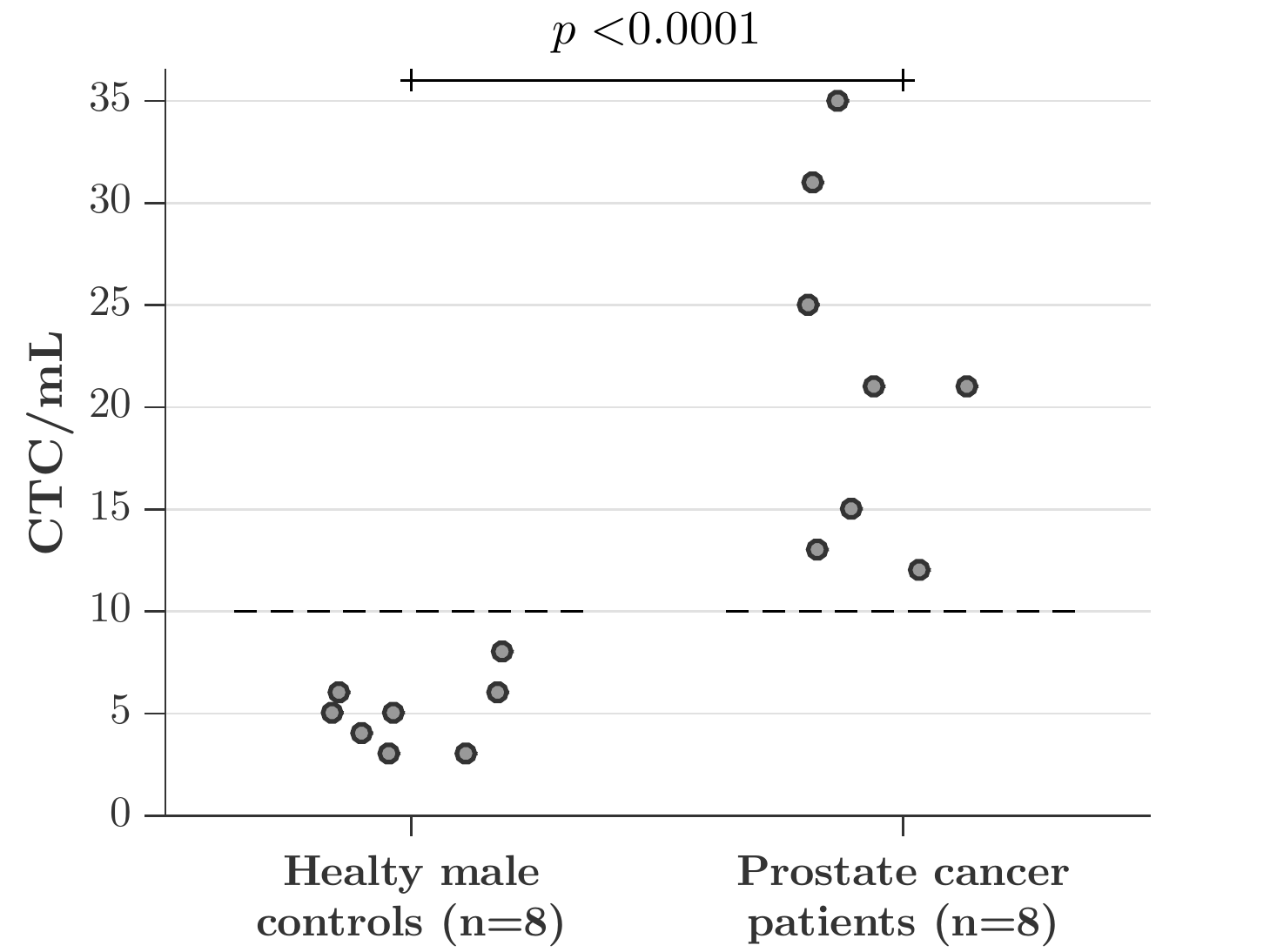}
\caption{CTC count obtained after processing blood samples from 8 healthy male donors and 8 metastatic prostate cancer patients. The number of events categorized as CTCs in control samples led us to establish a threshold of clinical significance of 10 CTC/mL. The differences between the number of CTCs found in these two groups were statistically significant ($p < 0.0001$).}
\label{fig:Graph}
\end{figure}

\begin{figure*}[t!]
\centering
\includegraphics[width=1\linewidth]{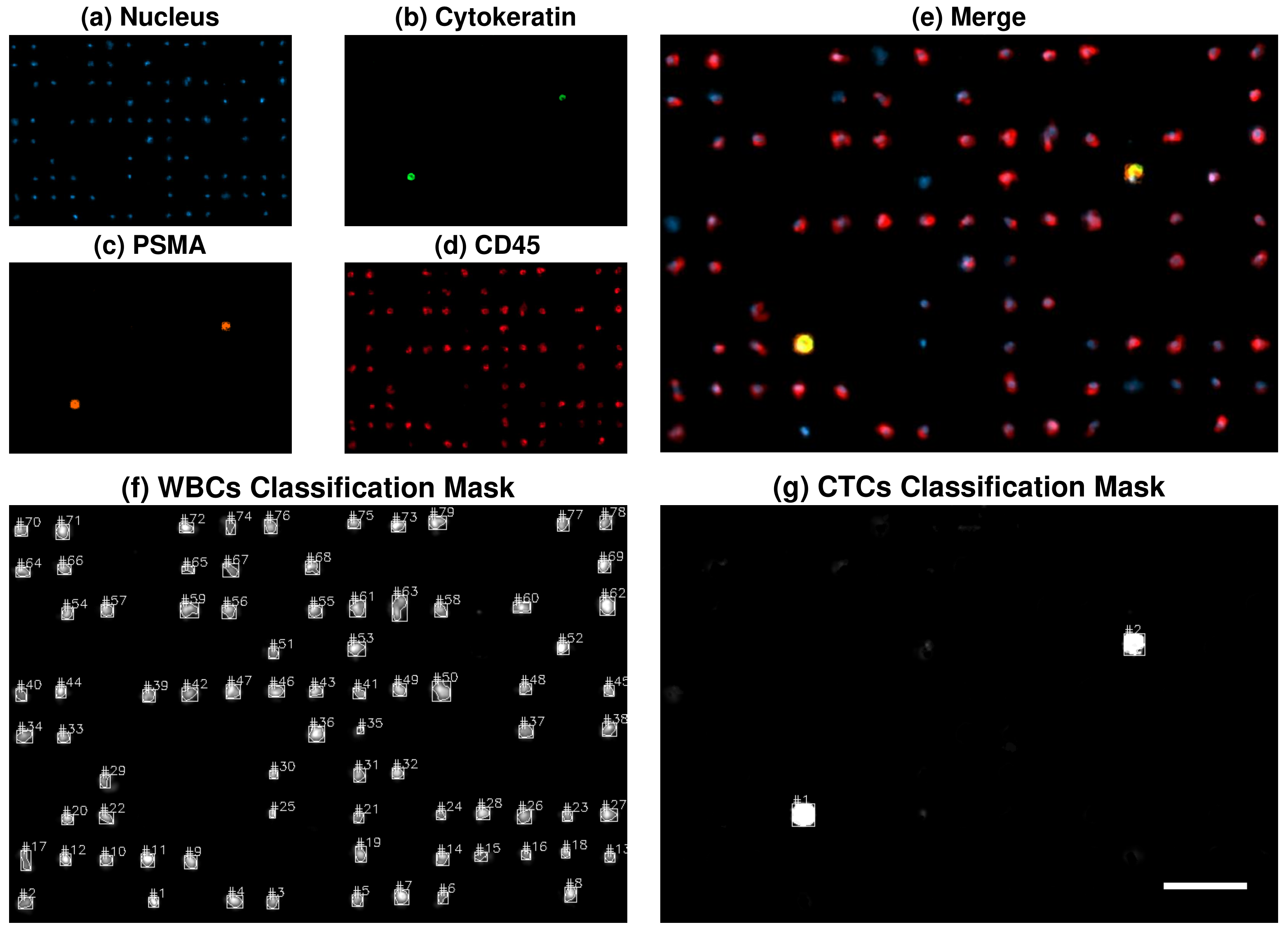}
\caption{Fluorescent micrograph of CTCs isolated from a metastatic cancer patient; (a) Nucleus, (b) Cytokeratin, (c) PSMA, (d) CD45 and (e) Merge. Classification masks generated for the identification of (f) WBCs and (g) CTCs. Scale bars of 50 $\mu$m.}
\label{fig:4CHSnap}
\end{figure*}

The number of cells categorized as CTCs, after analyzing control samples, ranged from 3 to 8 cells per mL of blood (mean $\pm$ SEM = 5 $\pm$ 0.597 CTCs/mL, median  = 5 CTC/mL). These results led us to establish a threshold of clinical significance of 10 CTC/mL. The CTC count in all samples from patients surpassed this value, ranging from 12 to 35 cells per mL of blood (mean $\pm$ SEM = 21 $\pm$ 2.957 CTCs/mL, median  = 21 CTC/mL). The difference between the means of both groups was statistically significant ($p < 0.0001$). Fig. \ref{fig:Graph} summarizes the results obtained after the analysis of patient and control samples. Fig. \ref{fig:4CHSnap} shows a micrograph of CTCs isolated from a patient diagnosed with metastatic prostate cancer, as well as the resulting classification masks generated by the counting software. 

Nucleated cells with a CK+/PSMA+/CD45+ phenotype were also observed in control and patient samples. The isolation of cells expressing epithelial, tissue-specific, and leukocyte markers has been reported in many instances and with different technologies \cite{stott2010a, lustberg2014}. Interestingly, these events were found in greater numbers than for the cells classified as tumoral and were more prevalent in patient samples than in controls; however, they were not counted as CTCs. 

\section*{Discussion}

CTCs have demonstrated their potential as a blood-based biomarker that can be used on a broad range of cancer-related clinical applications. This has been possible due to the recent development of several technologies, which have enabled the isolation and subsequent analysis of these malignant cells. However, despite the capture approach used, the non-scalable fabrication, prolonged sample processing times, and the lack of automation, of most of these technologies, represent substantial limitations that impede their transition from research tools to solutions employed in the standard clinical practice.

In this manuscript, we presented a novel automated microfiltration device that integrates an imaging system for the efficient isolation and rapid analysis of CTCs from blood samples. This platform can process 7.5 mL blood samples in less than 12 minutes, in comparison with other technologies that can only process significantly lower amounts of sample and require hours to do so \cite{nagrath2007, gupta2012, lu2013}. Due to the rarity of CTCs, technologies able to process high volumes of sample are desirable. Microfiltration devices are capable of processing samples in matter of minutes, however several of them have only reported to do it with small sample volumes \cite{hosokawa2010, hosokawa2013, tang2014, kang2015}. On the other hand, our device makes possible automatic on-membrane fixation and immunostaining without needing to disassemble the holder, enhancing practicality and avoiding cell loss during the staining process.

The integrated imaging system comprises a four-channel fluorescence microscope with motorized stage and an autofocus routine that was specifically designed for scanning the entire membrane with high precision. The relevance of the above relies on the fact that usually the filters suffer a deformation while processing the sample, resulting fundamental for the microscope to be capable of automatically focus when scanning large areas. Moreover, it also possesses a machine-vision algorithm that automatically counts the fluorescent events categorized as CTCs, thereby eliminating the subjective interpretation of operators and increasing the reproducibility of analysis, while decreasing the time needed to manually enumerate the CTCs captured. Further corroboration of classified cells can also be performed by a trained technician through a software interface that allows the individual visualization of cells selected as CTCs. These attributes facilitate the identification of predictive and therapeutic markers expressed in CTCs, such as AR-V7, HER-2, EGFR, among others. 

High capture efficiency and high reproducibility were achieved after processing spiked blood samples from healthy donors, with our device. Four different prostate cancer cells were used to characterize the device performance, and it was found that the capture efficiency of this platform was greater than 93\% among the four cells lines tested, having an average of 4144 non-specific nucleated events contaminating the enriched sample and small coefficients of variation, which denotes high reproducibility. Our recovery rates were superior than the ones obtained by several microfiltration devices \cite{xu2010, tan2010, riahi2014, kang2015, dewit2015, liu2019}. In addition, it was observed that the level of cellular contamination did not affect the identification of cells by immunostaining. Moreover, tumor cells remained viable after sample processing, demonstrating that the shear stress exerted on cells during filtering did not compromise their integrity. 

We were also able to successfully isolate total RNA from the cells captured on the filtration membrane. This was demonstrated by identifying the AR gene expression from 7.5 ml blood samples spiked with 50 LNCaP cancer cells (less than 7 cells per mL) by means of RT-PCR, as well as the point mutation T878A harbored in this cell line using Sanger sequencing. The relevance of this finding is that missense mutations often lead to the development of therapeutic resistance in several types of cancers. The noninvasive identification of these point mutations can aid physicians in monitoring changes in tumor genotypes during the course of treatments, enabling the personalization of cancer therapies. The results obtained after the reamplification of PCR products, derived from the cells captured on the membranes used to filter 7.5 mL samples spiked with a total of 15 cancer cells, suggest that the implementation of more sensitive techniques aimed at enriching target molecules, such as qRT-PCR and dPCR, could increase our limits of detection. Besides, due to the easy detachment of the membrane from the holder, single cells could be isolated by micropipette aspiration or by attaching a micromanipulator to the imaging system to perform single cell sequencing in order to enable the study of CTCs heterogeneity in cancer patients; unlike other filter-based microdevices that require additional steps to retrieve the tumor cells captured, which leads to further cell loss \cite{kang2015, xu2015}. 

In this work, blood samples from 8 patients diagnosed with prostate cancer were analyzed, as well as 8 control samples from healthy donors. Events classified as CTCs were observed in control samples, leading us to establish a threshold of clinical significance of 10 cancer cells per mL, similar to that reported by \cite{stott2010a}. CTCs above this threshold were detected in all patient samples, and their number was higher than those observed in control samples, even though some of these patients were clinically responding to the administered therapies. The number of CTCs captured by our device, points to the feasibility of monitoring the dynamic changes in CTC burden over time, implying that the treatment effectiveness follow-up would also be possible with our system. 

Over 80\% of the CTCs found in patient samples presented dual expression of cytokeratin and PSMA. Variability in the cytokeratin staining intensity, which is reported to occur during the EMT, was observed in CTCs, while PSMA expression remained consistent, supporting the use of this marker to discriminate between CTCs and blood cells in patients with advanced prostate cancer. Moreover, nucleated cells with a CK+/PSMA+/CD45+ phenotype were also observed in both the control and the patient samples. These cells were found in a greater concentration than the CTCs, and appear with a higher frequency in patient samples than in controls; these findings were similar to those reported when using other technologies \cite{stott2010a, lustberg2014}. Nevertheless, the clinical significance of these events warrants further study.

In conclusion, CTCs have demonstrated their potential as a powerful biomarker that can be continuously assessed to determine phenotypic and genotypic changes that confer therapeutic sensitivity/resistance during the course of cancer treatments. However, most of the technologies designed to capture these rare cells are not easily transferable to clinical practice. 
We have developed a novel membrane-based microfiltration device that integrates a fully automated sample processing unit and a machine-vision-enabled imaging system for the efficient isolation and rapid analysis of CTCs from blood; the platform allows the automation of the sample processing, immunostaining steps, analysis, and classification of fluorescent events for the identification of cancer cells. 
Capture efficiencies greater than 93\% and coefficients of variation below 7\% were achieved after processing samples from healthy donors spiked with different prostate cancer cell lines, while the isolated cells remained viable and suitable for molecular analysis. Moreover, CTCs above the established threshold of 10 cell/mL were detected in every sample from the 8 patients with metastatic disease.
Although large-scale clinical studies need to be done, these results, in addition to the fast processing time and large sample volume that can be processed by our platform, make this device a promising tool that can be rapidly integrated into clinical practice. 
Currently, a study with a larger cohort of prostate cancer patients is being carried out in order to continue the validation of this technology and the assessment of its clinical potential.

\bibliography{sample}

\section*{Acknowledgements}

JFYdL, MAE, CAA, LPVC, and AAB acknowledge the financial support received from CONACYT (Consejo Nacional de Ciencia y Tecnología, Mexico) through the FORDECYT project No. 273496. MMA and GTdS acknowledge the funding provided from CONACYT through the grant: Fronteras de la Ciencia-2442.

\section*{Author contributions statement}

JFYdL, BSG, DAH, JRDB, LPVC, and AAB conceived the experiments, JFYdL, BSG, DAH, JRDB, MAE, CAA, and JDWC conducted the experiments, MAE, CAA, JDWC, FC, JYLH, AMGT, JRYdL, and JLZM develop the system's hardware and software, MMA, GTdS, LSGG, CNSD, LPVC, and AAB facilitated equipment and reagents to conduct this work, LSGG and CNSD provided patient samples, JFYdL, BSG, DAH, JRDB, MAE, CAA, JDWC, MMA, GTdS, LSGG, CNSD, LPVC, and AAB analyzed and discussed the results obtained, JFYdL, BSG, DAH, JRDB, and CAA wrote the first draft of this manuscript. All the authors reviewed the final version of the manuscript.

\section*{Additional information}

\textbf{Competing Interests:}  

JFYdL, LPVC, and AAB are co-founders of Delee Corp., BSG, DAH, MAE, CAA, JDWC, FC, JYLH, AMGT, JRYdL, and JLZM are employees of Delee Corp. Delee Corp. funders had no role in the study design, data collection and analysis, decision to publish, or preparation of the manuscript. Other authors declare that there are no potential conflicts of interest regarding the authorship and/or publication of this article.

\end{document}